\documentclass[conference]{IEEEtran}
\usepackage{floatflt}
\usepackage{cite}

\usepackage{graphicx}
\usepackage{amsmath}
\usepackage{amssymb}

\usepackage{verbatim}

\usepackage{psfrag}

\usepackage{subfigure}
\usepackage{stfloats}
\usepackage{amsmath}
\usepackage{epstopdf}
\usepackage{algorithmic}
\newtheorem{definition}{Definition}

\newtheorem{example}{Example}

\newcommand{\mat}{\boldsymbol}

\begin{document}

\title{Energy Efficient Multiuser Scheduling:\\ Statistical Guarantees on Bursty Packet Loss}

\author{\authorblockN{M. Majid Butt\authorrefmark{1},~Eduard A. Jorswieck\authorrefmark{2} and Amr Mohamed\authorrefmark{1}}
\authorblockA{\authorrefmark{1}Department of Computer Science and Engineering, Qatar University, Qatar\\
Email: \{majid.butt,~amrm\}@qu.edu.qa}
\authorblockA{\authorrefmark{2}Department of Electrical Engineering and Information Technology,
Dresden University of Technology, Germany\\
Email: eduard.jorswieck@tu-dresden.de}
}
\maketitle

\begin{abstract}
In this paper, we consider energy efficient multiuser scheduling. Packet loss tolerance of the applications is exploited to minimize average system energy. There is a constraint on average packet drop rate and maximum number of packets dropped successively (bursty loss). A finite buffer size is assumed. We propose a scheme which schedules the users opportunistically according to the channel conditions, packet loss constraints and buffer size parameters. We assume imperfect channel state information at the transmitter side and analyze the scheme in large user limit using stochastic optimization techniques.

First, we optimize system energy for a fixed buffer size which results in a corresponding statistical guarantee on successive packet drop. Then, we determine the minimum buffer size to achieve a target (improved) energy efficiency for the same (or better) statistical guarantee. We show that buffer size can be traded effectively to achieve system energy efficiency for target statistical guarantees on packet loss parameters.
\end{abstract}
\begin{keywords}
Energy efficiency, opportunistic scheduling, stochastic optimization, simulated annealing.
\end{keywords}

\section{Introduction}
Radio resource allocation in wireless networks is becoming more complex due to stringent constraints imposed by Quality of Service (QoS) requirements.
The QoS requirements vary from application to application in terms of average/maximum packet delay and average packet loss rate. Energy efficiency is another important factor in today's
wireless networks design due to increasing cost of network operation. The radio resource algorithms should aim at utilizing every soft
QoS requirement to improve the energy efficiency of the network, e.g., the authors in \cite{Zhang:2013} investigate power-delay tradeoff and propose
resource allocation schemes to minimize power consumption subject
to a delay QoS constraint where the delay constraint is in terms of queue-length decay rate. Energy-performance tradeoffs have been addressed in
different settings in \cite{majid_WCL12,Neely}.
Most of the work focuses on exploiting delay tolerance to optimize the system energy. However, data loss tolerance is another aspect that
can be exploited to save system energy.

We consider a multiple access system where the users have soft average packet drop constraint and a hard constraint on
the maximum number of packet allowed to be dropped \emph{successively}. This constraint is referred to as continuity constraint \cite{majid_TWC:13}. The work in \cite{majid_TWC:13,majid_PIMRC_13} takes this constraint into account for packet scheduling. The idea is to allow
intentional (but bounded) packet drop to save energy if application can tolerate it without a major deterioration in quality of experience. The authors
assume a perfect channel state information (CSI) on the transmitter and receiver
sides. A channel threshold based scheduling scheme is introduced and analyzed. However, perfect CSI condition is hard to achieve in practical networks
due to limitation on feedback information. This paper extends the work for imperfect CSI case and addresses the resulting effects on the
system energy and scheduling process.
The effect of imperfect channel information is modeled through a corresponding packet drop probability.

We have two reasons for packet drop in our problem settings:
\begin{enumerate}
  \item {Intentional packet drop at the transmitter depending on the application loss tolerance to save energy if applications's loss tolerance
      permits.}
  \item {Packet drop due to imperfect CSI at the transmitter side which implies that channel state is worse than the estimated one and results in
      packet loss after
      transmission.}
\end{enumerate}
The scheduling algorithm design for the packet loss tolerant application should take the packet loss due to imperfect CSI into account and adapt its intentional packet drop rate accordingly to maintain a bounded average packet drop rate.
The main contribution of this work is to evaluate this
effect through packet level channel model. Moreover, we extend the results to the (more practical) case of statistical guarantees on continuity
constraint as compared to hard guarantees in \cite{majid_PIMRC_13}.

The rest of the paper is structured as follows. Section \ref{sec:system_model} introduces the system model and preliminaries. We model the proposed
scheme using Markov chain in Section \ref{sect:Scheme}. The optimization problem is formulated in Section \ref{sect:optimization}. We evaluate the
scheme numerically in Section \ref{sect:results} and conclude with the main contributions in Section \ref{sect:conclusions}.

\section{System Model and Preliminaries}
\label{sec:system_model}
We follow the system model used in \cite{Ralf1,majid_TWC:13} and consider a multiple-access system with $K$ users randomly placed within a certain
area. Every scheduled user requires an average rate $R_k=\frac{C}{K}$ where $C$ denotes the system spectral efficiency.

\subsection{Propagation Channel Model}
We consider an uplink scenario where time is slotted such that each user $k$ experiences a channel gain $h_k(t)$ in a
time slot $t$. The channel
gain $h_k(t)$ comprises of path loss component $s_k$ and small-scale
fading $f_k(t)$ such that $h_k(t)=s_kf_k(t)$. The path loss is a function of
the distance between the transmitter and the receiver and remains constant within the time scale considered in this work. Small-scale fading depends
on
the scattering environment. It changes
from slot to slot for every user and is independent and identically
distributed (i.i.d) across both users and slots, but remains constant during the time span of a single time slot. The multi-access channel is
described
by the input (X) and output (Y) relation as
\begin{equation}
Y_k(t)=\sum_{k=1}^K \sqrt{h_k(t)}X_k(t)+ Z(t)
\end{equation}
where $Z$ represents additive i.i.d. complex Gaussian
random variable with zero mean and unit variance. The distribution of $h_k(t)$ differs from user to user.
\subsection{Packet level channel Model}
\label{sec:channel_model}
The CSI is assumed to be known at the transmitter, but it is not perfect. As a result of imperfect CSI, the transmitter is not able to compute correct power level for the assigned rate. This could result in packet loss. We model the effect of imperfect CSI by a probability
$\nu_d$ that a resulting
transmission is not successful. Furthermore, we assume if the estimated channel was not good enough to support the rate, all the packets transmitted in
a single transmission are lost. The information about the dropping of the packet(s) is conveyed by the receiver to the transmitter through a perfect
feedback channel.
This is termed as packet level channel modeling in the literature.
We assume that the feedback arrives at the transmitter by the next scheduling instance. The unsuccessfully transmitted packet(s) is buffered for (possible) retransmission if the buffer has the capacity to store it for the next time slot, dropped otherwise.

\subsection{Statistical Guarantees on Continuity Constraint}
The model considered in \cite{majid_PIMRC_13} assumes that continuity constraint can be met with probability one. It is not practicable to assume
that a packet can be transmitted with probability one when $N$ packets have been dropped successively; where $N$ is termed as continuity constraint parameter. We extend our framework in the direction of providing
statistical guarantees on continuity constraint, i.e., a user violates the continuity constraint with a probability $\gamma$. If channel conditions are
not good after dropping $N$ packets successively, the user is still allowed to drop a finite amount of packets corresponding to $\gamma$. We define the event
of violation of continuity constraint as the number of time slots a packet is dropped after
successively dropping $N$ packets already.


We allow multiple users to be scheduled in a single time slot to minimize $\gamma$. If only a single user is scheduled
per time slot, all the users other than the scheduled one may have to drop the packets (intentionally) which results in increase in $\gamma$ rapidly.
We have no control over the packets dropped due to channel impairments, but packet scheduler can be designed such that
$\gamma$ is bounded by facilitating maximum scheduling of the users who already have dropped $N$ packets successively.

The analysis of the scheme is based on asymptotic
user case which implies that there is no limit on the number of
users scheduled simultaneously. We use superposition coding and successive interference cancelation (SIC) mechanism for successful transmission of
data
streams of simultaneously scheduled users. Let ${\mathcal{K}}$ denote the set of users
to be scheduled and $\Phi_k$ be the
permutation of the scheduled user indices that sorts the channel gains in increasing order, i.e.\
$h_{\Phi_1}\le \cdots \le h_{\Phi_k}\le \cdots \le
h_{\Phi_{|{\mathcal K}|}}$. Then, the energy of the scheduled user
$\Phi_k$ with rate $R_{\Phi_k}$, is given by
\cite{Tse3,Ralf1}
\begin{equation}\label{eqn:power}
 E_{\Phi_k} = \frac{Z_0}{h_{\Phi_k}} \left({2^{\sum_{i\leq k}R_{\Phi_i}}-2^{\sum_{i<k}R_{\Phi_i}}}\right).
\end{equation}
where $Z_0$ denotes the noise power spectral density.
\section{Analysis of the Scheduling Scheme}
\label{sect:Scheme}
In this section, we briefly review the scheduling scheme presented in \cite{majid_PIMRC_13} for perfect CSI scenario. Then, we model the scheme for
the imperfect CSI case and analyze the effect on the scheduling decisions and average system energy.
\subsection{Review of the Scheme with Perfect CSI}
\label{subsect:review}
We consider a constant arrival of a single packet in the buffer of each user in each time slot for simplicity. However, this model can be extended to
the random arrival case where multiple packets arriving in the same time slot are treated as a single packet \cite{majid_TWC:13}. Every arriving packet
is queued in the user's buffer after arrival.

The design of the scheme is based on the asymptotic case when $K\to \infty$. In this case, multiuser scheduling problem can be broken into a single
user scheduling problem such that every user takes the scheduling decision independent of the other users \cite{majid_TWC:13}.
The scheduling decisions for every user in each time slot are based on the instantaneous channel condition and the scheduling thresholds; which are
optimized by taking into consideration the continuity constraint parameter $N$, maximum buffer size $B$, average packet dropping probability
$\theta_{\rm tar}$ and user's small scale fading distribution. The number of thresholds equals the number of buffered packets and the scheduler decides
how many packets are scheduled in a single time slot based on the channel conditions. If no packet is scheduled, all the packets (including the
recently arrived packet) are buffered if
buffer size allows. If the buffer is full, the oldest packet in the buffer is dropped. When the user has dropped $N$ packets successively, the head of
line (HOL) packet has to be scheduled regardless of the channel conditions.
\subsection{Finite State Markov Chain Model}
\label{sect:FSMC}
We extend the results for the scheduling scheme proposed in \cite{majid_PIMRC_13} to the case of imperfect CSI. As explained in Section
\ref{sec:channel_model}, the effect of imperfect CSI is modeled through packet level description such that $\nu_d$ denotes packet drop probability and
$\nu_s=1-\nu_d$ is the probability of a successful transmission.

We model the scheme discussed in Sec. \ref{subsect:review} using a finite state Markov chain (FSMC). Let $i\leq B$ and $j\leq N$ denote the number of
packets buffered and dropped successively at time $t$. Then, state $p$ at time $t$ is defined by the summation $i+j$. At
the start of the process, $p$ equals zero. If a packet is not scheduled, it is buffered and $i=1$
(while $j=0$), thereby the system makes transition to next state $q=1$. Remember $p(t+1)=q(t)$ in FSMC. When the buffer is full, an event of not
scheduling a packet results in a packet drop, thereby $j$ starts increasing and $i=B$ remains fixed until there is a room in the buffer for unscheduled
packets due to scheduling of previously buffered packets. The event of dropping/buffering of the packet results in forward state transition to next
state $q=p+1$. The size of FSMC is determined by the buffer size and continuity constraint parameters such that $M=B+N$.

We did not consider the event of packet drop due to imperfect CSI in the state space description yet. As we assume that feedback for the
successful/unsuccessful transmission (ACK/NACK) arrives by the end of time slot, the system buffers the scheduled packet(s) by the end of time slot. If the transmitter receives an ACK, the packets are dropped from the buffer as they have been received successfully. In case of a NACK, the buffered packets are treated in the same way as intentional packet dropping, i.e. buffer if there is a room or drop otherwise. In contrast to the case of ACK, dropping a packet in case of a NACK occurs solely due to insufficient buffer capacity and affects system performance similar to intentional packet drop scenario. The packet drop due to imperfect CSI needs to be modeled in the system separately due to its different effect on system energy. Intentional packet dropping (without transmission) does not cost any energy to the system while packets dropped due to imperfect CSI result in waste of energy without transmitting any data successfully.

In an FSMC model, we define $\alpha_{pq}$ as
\begin{eqnarray}
\alpha_{pq} = {\rm Pr}(S_{t+1}=q|S_t=p)
=\begin{cases}
\hat{\alpha}_{pq} & \forall p,q\leq \min(p,B)\\
\tilde{\alpha}_{pq} & \forall p,q=p+1\\
0& \mbox{else}
\end{cases}
\end{eqnarray}
where
\begin{eqnarray}
\alpha_{pq} &=& \mbox{Transition probability from state $p$ to $q$.}\nonumber\\
\hat{\alpha}_{pq} &=& \mbox{Transition probability from state $p$ to $q$}\nonumber\\&& \mbox{when scheduling of one or more packets
occurs.}\nonumber\\
\tilde{\alpha}_{pq} &=& \mbox{Transition probability from state $p$ to $q$}\nonumber\\&& \mbox{when no packet is scheduled.}\nonumber
\end{eqnarray}
$\alpha_{pq}$ is a function of $\hat{\alpha}_{pq}$ and $\tilde{\alpha}_{pq}$.

To define $\hat{\alpha}_{pq}$ and $\tilde{\alpha}_{pq}$, we define a scheduling threshold.
\begin{definition}[Scheduling Threshold $\kappa_{pq}$] It is defined as the minimum small scale fading value $f$ required to make a state transition
from state $p$ to $q$ such that
\begin{equation}
\hat{\alpha}_{pq} =
{\rm Pr}\bigl(\kappa_{pq}<f\leq\kappa_{p(q-1)}\bigr)\quad 0\leq q\leq \min(p,B).
\label{eqn:alpha1}
\end{equation}
where $\kappa_{p0^-}$ is defined to be infinity with $S_{0-}$ denoting a dummy state before $S_0$.
\end{definition}
From scheduling point of view, it is advantageous to schedule more packets for good fading states. Therefore, the scheduling thresholds quantize the
fading vector to optimize the number of scheduled packets according to the fading.
In a state $p\geq q$, the scheduler makes a state transition to state $q$ such that \cite{majid_PIMRC_13}
\begin{equation}
q =  \arg\min_{\acute{q}} \kappa_{p\acute{q}}<f\leq\kappa_{p(\acute{q}-1)}\quad 0\leq \acute{q}\leq \min(p,B).
\label{eqn:decision}
\end{equation}
For a state transition from state $p$ to $q$, the number of the scheduled packets is given by
\begin{equation}
L(p,f) = \min(p,B)-q+1,
\label{eqn:packets}
\end{equation}
where $q$ is determined uniquely by (\ref{eqn:decision}). Note that the number of scheduled packets cannot exceed $\min(p,B)$ because of finite
capacity of buffer. We denote $\min(p,B)$ by $\mu$ in the rest of this article for convenience.
The probability of not scheduling any packet for transmission is given by
\begin{eqnarray}
\tilde{\alpha}_{pq}&=& {\rm Pr} (f\leq\kappa_{p\mu}),\qquad 0\leq p< M,q=p+1\\
&=&1-\sum_{q=0}^\mu\hat{\alpha}_{pq}~.
\label{eqn:alpha2}
\end{eqnarray}
where $\kappa_{p\mu}$ denotes the minimum thresholds to schedule at least one packet.

\subsection{Modeling $\gamma$ in FSMC}
Ideally one would like to schedule a packet with probability one when $p=M$.
As explained earlier, it is not practical to apply 'water filling' principle on any arbitrary channel due to power limitations of the transmitter.
Thus, a packet is not scheduled if fading is worse than a minimum value even in state $M$ which contributes to $\gamma$ in addition to packet
dropping due to imperfect CSI. To handle the event of unscheduled or/and lost HOL packet in state $M$, we define a self transition $\alpha_{MM}$ where no packet is scheduled in contrast to other self state transitions.

According to our FSMC model,
\begin{eqnarray}
\gamma&=&\alpha_{MM}\pi_M= \Bigl(\tilde{\alpha}_{MM}+\nu_d \sum_{q=0}^{B}\hat{\alpha}_{Mq}\Bigr)\pi_M\\
&=& \Bigl(1-\nu_s \sum_{q=0}^{B}\hat{\alpha}_{Mq}\Bigr)\pi_M
\label{eqn:outage}
\end{eqnarray}
where $\pi_M$ is steady state transition probability for state $M$.
Thus, $\alpha_{pq}$ for any states $p,q$ is modified as
\begin{eqnarray}
\alpha_{pq}&=&\nu_s\hat{\alpha}_{pq},\quad 0\leq p\leq M,0\leq q\leq \mu\\
\alpha_{pq}&=& \tilde{\alpha}_{pq} + \nu_d\sum_{m=0}^\mu\hat{\alpha}_{pm},\quad 0\leq p< M,q=p+1
\end{eqnarray}
\begin{example}
Let us explain FSMC model with the help of an example with $B=2$, $N=1$ as in Fig. \ref{fig:state diagram}.
\begin{figure}
   \includegraphics[width=3.2in]{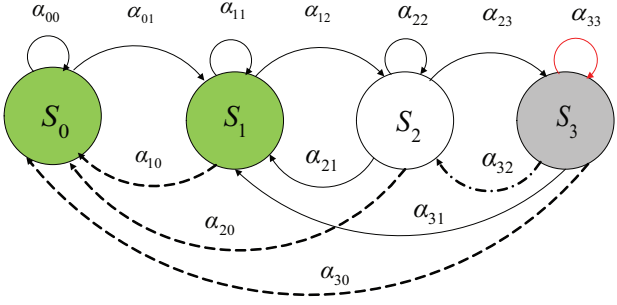}
  \caption{State transition diagram of the scheme for the case $B=2, N=1$. $\alpha_{MM}$ represents state transition probability related to $\gamma$.
  }
  \label{fig:state diagram}
\end{figure}
For this example, we evaluate the transition probability matrix $\mat{Q}$.

For state 0, $\alpha_{00}$ is the probability that a packet is scheduled for transmission and received successfully.
\begin{equation}
\alpha_{00} = \nu_s\hat{\alpha}_{00}
\end{equation}
Transition from state 0 to 1 is the result of un-scheduled; and scheduled but unsuccessful transmission events.
Thus,
\begin{equation}
\alpha_{01} = \tilde{\alpha}_{01} + \nu_d\hat{\alpha}_{00}
\end{equation}
Similarly, for state 1,
\begin{eqnarray}
\alpha_{1q} &=&\nu_s\hat{\alpha}_{1q}\qquad q=0,1\\
\alpha_{12}&=& \tilde{\alpha}_{12} + \nu_d(\hat{\alpha}_{10}+\hat{\alpha}_{11})
\end{eqnarray}
Following the same line of arguments, the matrix $\mat{Q}$ can be written as a summation of two matrices such that
\begin{equation}
\mat{Q}=\mat{Q_{\rm s}}+\mat{Q_{\rm c}}
\end{equation}
where
\begin{equation}
\mat{Q} = \left( \begin{array}{cccc}
\alpha_{00} & \alpha_{01}&0  &0 \\
\alpha_{10} & \alpha_{11} & \alpha_{12} &0 \\
\alpha_{20} & \alpha_{21} & \alpha_{22} &\alpha_{23} \\
\alpha_{30}& \alpha_{31}& \alpha_{32} &\alpha_{33}\end{array} \right) \label{eqn:Drop_STM}
\end{equation}
\begin{equation}
\mat{Q}_{\rm s} = \left( \begin{array}{cccc}
\nu_s\hat{\alpha}_{00} & \tilde{\alpha}_{01}&0  &0 \\
\nu_s\hat{\alpha}_{10} & \nu_s\hat{\alpha}_{11} & \tilde{\alpha}_{12} &0 \\
\nu_s\hat{\alpha}_{20} & \nu_s\hat{\alpha}_{21} & \nu_s\hat{\alpha}_{22} &\tilde{\alpha}_{23} \\
\nu_s\hat{\alpha}_{30}& \nu_s\hat{\alpha}_{31}& \nu_s\hat{\alpha}_{32} &\tilde{\alpha}_{33}\end{array} \right) \label{eqn:Drop_STM1}
\end{equation}
and
\begin{equation}
\mat{Q}_{\rm c} = \nu_d \left( \begin{array}{cccc}
0 & \sum_{q=0}^0 \hat{\alpha}_{0q}&0  &0 \\
0 & 0 & \sum_{q=0}^1 \hat{\alpha}_{1q} &0 \\
0 & 0& 0 &\sum_{q=0}^2 \hat{\alpha}_{2q} \\
0 & 0& 0&\sum_{q=0}^2 \hat{\alpha}_{3q}\end{array} \right) \label{eqn:Drop_STM2}.
\end{equation}
$\mat{Q}_{\rm c}$ captures the effect of imperfect CSI while $\mat{Q}_{\rm s}$ is optimized scheduling decision matrix. Note that this model implies
that it
is not possible to achieve continuity constraint with probability one if $\nu_d>0$ and only statistical guarantees can be provided.
\end{example}
\section{Mathematical Formulation of the Problem}
\label{sect:optimization}
The objective of the optimization problem is to minimize the system energy for a soft average packet drop rate constraint and statistical guarantee on
continuity constraint. We formulate the optimization problem using the FSMC model.
Each scheduled packet is treated as a virtual user for the analysis purpose.
The average system energy per transmitted information bit at
the large system limit $K \to \infty$ is given by \cite{Ralf1}
\begin{equation}
\label{eqn:energy_function}
\frac{E_{\rm b}}{N_0}=  \log(2) \int\limits_0^\infty \frac {2^{C \,{\rm P}_{h,\rm
VU}(x)}}x\, {\rm dP}_{h,\rm VU}(x)
\end{equation}
where ${\rm P}_{h,\rm VU}(\cdot)$ denotes the cumulative distribution function (cdf) of the fading of the scheduled virtual users (VUs). In the large system limit, the
state transitions depend only on the small scale fading distribution as the path loss for VUs follows the same distribution as the path loss of the
users.
Thus, the optimization problem\footnote{This section follows the work in \cite{majid_PIMRC_13} closely, but developments of the next section require its inclusion for completeness and clarity of the discussion.} is formulated as
\begin{eqnarray}
\label{eqn:objective}
&\min_{\mat{Q}\in \Omega} \frac{E_{\rm b}}{N_0}&\\
&\mbox{s.t.}:\begin{cases}\mathcal{C}_1: 0\leq\sum_{m=0}^{\mu}\alpha_{pm}\leq 1 &   0 \leq \alpha_{pm} \leq 1,\\&0\leq p\leq M\\
\mathcal{C}_2: \theta_r\leq \theta_{\rm tar} &  \mat{Q}\in \Omega\\
\mathcal{C}_3:\sum_{q=0}^M \alpha_{pq}=1& 0\leq p \leq M\\
 \mathcal{C}_4:B+N=M & B<\infty, N<\infty
\end{cases}&
\label{eqn:optimization}
\end{eqnarray}
where $\Omega$ denotes the set of permissible matrices for $\mat{Q}$ and $\theta_r$ is the average packet drop rate for a fixed $\mat{Q}$ and
given by
\begin{eqnarray}
\theta_r &=& \sum_{p=B}^{M-1}\alpha_{p(p+1)}\pi_p+\alpha_{MM}\pi_M\\
&=& \sum_{p=B}^{M}\bigl(1-\nu_s\sum_{m=0}^{B}\hat{\alpha}_{pm}\bigr)\pi_p~.
\label{eqn:drop_cons2}
\end{eqnarray}
Equation (\ref{eqn:drop_cons2}) is a result of combining $\mathcal{C}_1$ and $\mathcal{C}_3$ in (\ref{eqn:optimization}).
The forward transition for the states $B\leq p< M$ and self state transition $\alpha_{MM}$ represent the events of packet drop and the summation
over the corresponding transition probabilities gives the average dropping probability in (\ref{eqn:drop_cons2}). The summation starts from state $B$
as the
unscheduled packets are buffered for $p<B$. For a fixed $p$, the corresponding channel-dependent optimal scheduling thresholds can be computed from the
optimized $\vec{\alpha_p^*}=[\alpha_{p0}^*,\dots \alpha_{p\mu}^*]$ using (\ref{eqn:alpha1}).
The violation probability on continuity constraint $\gamma$ for fixed $B$ and $N$ parameters is computed from $Q^*$ using (\ref{eqn:outage}). Let us denote $\gamma$ for this special case by $\gamma_m$ where the \emph{maximum} energy efficiency can be achieved for fixed $B,N,\theta_{\rm tar}$ parameters and relaxing $\gamma$ further does not help to improve energy efficiency due to coupling of $\gamma$ with $N$ and $\theta_{\rm tar}$.

If the statistical guarantees has to be improved further, we apply an upper bound on $\gamma$ such that $\gamma\leq \epsilon$ where $\epsilon$ is a small constant representing the target statistical guarantee. This constraint appears as an additional constraint in (\ref{eqn:optimization}) such that
\begin{eqnarray}
      \mathcal{C}_5: \gamma\leq \epsilon\quad 0\leq\epsilon\leq \theta_{\rm tar}
      \label{eqn:gamma_constraint}
\end{eqnarray}
because $\theta_r = \sum_{p=B}^{M-1} \alpha_{p(p+1)} \pi_p + \gamma$. Consequently, the improved $\gamma$ is achieved at the increased energy cost. Theoretically $\epsilon$ is upper bounded by $\theta_{\rm tar}$; but $\gamma_m$ upper bounds $\epsilon$ (tightly) at a
value lower than $\theta_{\rm tar}$ due to the reasons explained above.

The probability density function (pdf) of the small scale fading of the
scheduled VUs is given by
\begin{equation}
{\rm p}_{f,\rm VU} (y) = \sum\limits_{p=0}^M c_p\pi_pL(p,y)\,{\rm p}_{f}(y) \label{eqn:SVU_fading}
\end{equation}
where ${\rm p}_{f}(y)$ and $c_p$ denote the
small scale fading distribution and a normalization constant respectively while $L(p,y)$ is given by (\ref{eqn:packets}).
The derivation of the cdf of VUs can be found in \cite{majid_PIMRC_13}.
The channel distribution for the scheduled VUs can be computed using fading and the
path loss distributions which in turns is used to compute system energy in (\ref{eqn:energy_function}).

\subsection{Trading Buffer for Improved Guarantees on $\gamma$}
\label{subsec:optimizations}
We would like to achieve $\epsilon\leq\gamma_m$ at improved energy by increasing the value of $B$ for a fixed $N$. Let us denote the optimal solution of the programming problem in previous section by $ Q^*(B,\theta_{\rm tar},\epsilon)$ as a function of $B$, $\theta_{\rm tar}$ and target violation probability on continuity constraint $\epsilon$. Let $\frac{E_b}{N_0}(Q^*(B,\theta_{\rm tar},\epsilon))$ be the corresponding system energy and $\Delta E$ represents the target energy gain. Now, the optimization is performed
over $B\in \Phi$ where $\Phi$ is a set of possible buffer sizes. For every candidate $B\in\Phi$, optimization in (\ref{eqn:objective}) and
(\ref{eqn:optimization}) is performed again with inclusion of $\mathcal{C}_5$. The aim of the optimization is to find minimum value of $B$ which gives energy less than $\Big(\frac{E_b}{N_0}(Q^*(B,\theta_{\rm tar},\epsilon))-\Delta E\Big)$ at $\epsilon$:
  \begin{eqnarray}
  	&{\rm Find}~  B^* \in \Phi \quad {\rm s.t.} \quad \gamma(Q^*(B^*,\theta_{\rm tar})) \leq \epsilon \quad {\rm and}&  \\
  	 &\frac{E_b}{N_0}(Q^*(B^*,\theta_{\rm tar},\epsilon)) - \frac{E_b}{N_0}(Q^*(B,\theta_{\rm tar},\epsilon)) \geq \Delta E ,& ~   B \in \Phi   \nonumber
  \label{eq:form1}
  \end{eqnarray}
The suitable value of $B$ is highly dependent on the application. For example, wireless sensor networks would prefer large $B$ due to battery
requirements whereas multimedia applications prefer small $B$ due to stringent delay requirements on data delivery.


\subsection{Stochastic Optimization}
The optimization problem formulated in (\ref{eqn:objective}) and (\ref{eqn:optimization}) is not convex and belongs to class of problems called
stochastic optimization problems. There are a few heuristic techniques in literature to solve such problems like genetic algorithm, Q-learning, neural
networks, etc. We use Simulated Annealing (SA) algorithm to solve the problem. As the name suggests, the algorithm originates from statistical
mechanics area and has been found quite useful to solve different combinatorial optimization problems like traveling salesman.

In SA algorithm, a random configuration in terms of transition probability matrix $\mat{Q}$ is presented in each step and system energy as an objective
function is evaluated only if $\mat{Q}$ fulfills all the constraints in (\ref{eqn:optimization}). If system energy improves the previous best solution,
the candidate configuration is selected as the best available solution. However, a candidate configuration can be treated as best solution with a
certain temperature dependent probability even if the new solution is worse than the best known solution. This step is called \emph{muting} and helps
the system to avoid local minima. The muting step occurs frequently at the start of the process as temperature is selected very high and decrease as
temperature is decreased gradually. Thus, the term temperature determines the rate of muting process. In literature, different cooling temperature
schedules have been employed according to the problem requirements. In this work, we employ the following cooling schedule, called fast annealing (FA)
\cite{FA}.
In FA, it is sufficient to decrease the temperature linearly in each step $b$ such that,
\begin{equation}
\label{eqn:BA} T_b = \frac{T_0}{c_{\rm sa}*b+1}
\end{equation}
where $T_0$ is a suitable starting temperature and $c_{\rm sa}$ is a constant which depends on the requirements of the problem.
We skip details of the SA scheme due to space limitations. The interested reader is referred to \cite{SA_1} for details of the algorithm.

\section{Numerical Results}
\label{sect:results}
We assume that the users are placed uniformly at random
in a circular cell except for a forbidden region around the access point of
radius $\delta=0.01$. The path loss exponent equals 2 and the
path loss distribution follows the model in \cite{Ralf1}. All the users experience independent small-scale fading with
exponential distribution with mean one. Spectral efficiency is $0.5$ bits/s/Hz for all simulations. In SA algorithm, $100$ temperature values are
simulated according to FA temperature schedule while $50(M+1)$ random configuration of transition probability matrix are generated for a single
temperature iteration.

To compute $\gamma_m$, we perform optimization in (\ref{eqn:optimization}) without applying constraint in
(\ref{eqn:gamma_constraint}) and the \emph{best}\footnote{We avoid using term energy optimal here as SA is a heuristic algorithm and solution cannot be proven optimal.} solution matrix $Q^*$ is obtained. The value of $\gamma$ computed via (\ref{eqn:outage}) for $Q^*$ gives us $\gamma_m$ and upper
bounds $\epsilon$. Table \ref{table_unconstrained} shows numerical values of system energy and $\gamma_m$ for different $N$ and fixed $B=0,\theta_{\rm
tar}=0.3$ values while $\nu_d$ equals $0.02$.
Based on numerical results in Table \ref{table_unconstrained}, we evaluate the tradeoffs addressed in Section \ref{subsec:optimizations}.

\begin{figure*}
\centering
  \subfigure[System energy as a function of $\epsilon$ for a system with fixed $B=0$.]{\includegraphics[width=3.5in]{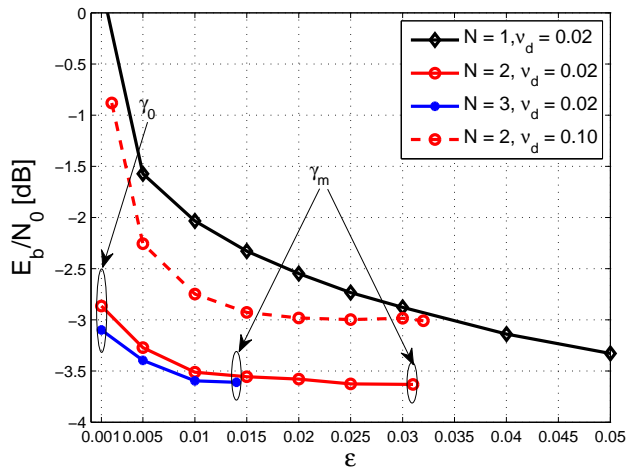}
 \label{fig:energy_nobuffer}}
 \subfigure[Achieved packet drop rate $\theta_r$ from (\ref{eqn:drop_cons2}) as a function of $\epsilon$ for the same parameters as in Fig.
 \ref{fig:energy_nobuffer}.]{\includegraphics[width=3.5in]{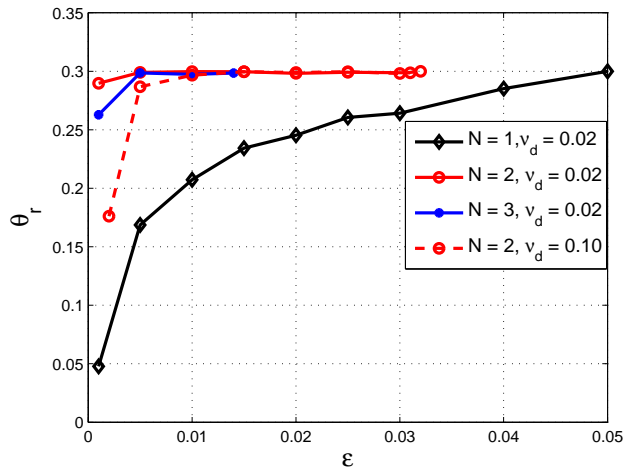}
  \label{fig:drop_nobuffer}}
\subfigure[System energy as a function of $\epsilon$ when $B>0$. All other parameters are the same as in Fig.
\ref{fig:energy_nobuffer}]{\includegraphics[width=3.5in]{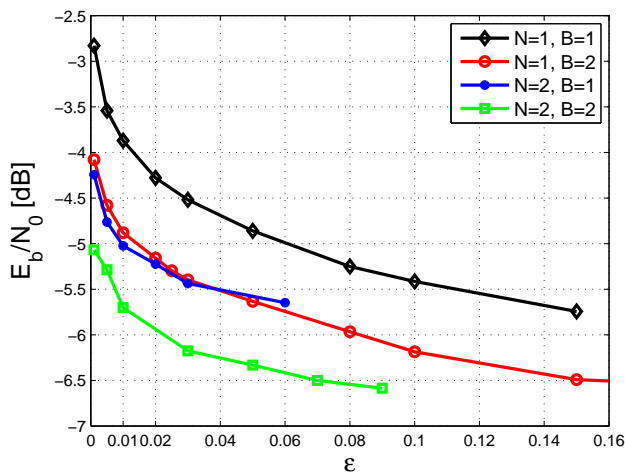}
  \label{fig:energy_buffer}}
 \caption{System energy and packet drop behavior as a function of $\epsilon$. $\theta_{\rm tar}$ is fixed to $0.3$ for all simulations.}
 \label{fig:energy_vs_gamma}
\end{figure*}

Fig.~\ref{fig:energy_vs_gamma} demonstrates the effect of imposing constraint $\epsilon\leq \gamma_m$ on system performance when $\theta_{\rm tar}=0.3$. We
evaluate $\mathcal{C}_5$ alongwith $\mathcal{C}_1-\mathcal{C}_4$ in (\ref{eqn:optimization}) for the candidate $\mat{Q}$ before evaluation of (\ref{eqn:energy_function}) in SA algorithm. We observe in
Fig.~\ref{fig:energy_nobuffer} that decreasing $\epsilon$ has an associated energy cost and the solution becomes suboptimal by energy point of view.
Also, $\gamma$ can never approach zero as long as $\nu_d>0$ and packet dropping due to imperfect CSI cannot be completely eliminated. For a given
set of parameters and fixed $\nu_d$, the minimum value of achievable $\epsilon$ is denoted by $\gamma_0$ which lower bounds $\epsilon$ such that
$\gamma_0\leq\epsilon\leq\gamma_m$. The greater the value of $\nu_d$, the greater is $\gamma_0$. For instance, increasing $\nu_d$ from $0.02$ to $0.1$
for the case $N=2$ raises $\gamma_m$ from $0.001$ to $0.002$ while system energy increases for all values of $\epsilon$ as well. We observe that bounds
on $\epsilon$ (in the form of $\gamma_0$ and $\gamma_m$) become tighter as $N$ increases for the fixed $\theta_{\rm tar}$. This is due to the fact that
allowing large $N$ increases degrees of freedom (DoF) for the system and the effect of parameter $\epsilon$ on system energy is minimized.

\begin{table}
\center
\renewcommand{\arraystretch}{1.0}
\caption{$\gamma_m$ and System Energy}
\label{table_unconstrained}
\begin{tabular}{|c||c||c|}
\hline
$N$ & $\gamma_m$&$E_b/N_0$\\
\hline
1 &0.09 &-3.63 dB\\
2&0.032&-3.63 dB\\
3&0.014&-3.61 dB\\
 \hline
\end{tabular}
\end{table}

Correspondingly, Fig.~\ref{fig:drop_nobuffer} demonstrates that achieved average packet drop rate $\theta_r$ (calculated via (\ref{eqn:drop_cons2}))
approaches $\theta_{\rm tar}$ for large $\epsilon$ and remains almost identical thereafter. This implies that all the extra energy cost is contributed
by strict guarantees on continuity constraint. When $\epsilon$ is very small, the energy optimal $Q^*$ provides a $\theta_r$ which is much less that
$\theta_{\rm tar}$ and severely sub optimal. We conclude that a strict statistical guarantee on continuity constraint has a severe plenty in terms of energy and even other DoF
(like relaxed $\theta_{\rm tar}$) cannot be utilized efficiently.


Fig.~\ref{fig:energy_buffer} demonstrates the energy gain achieved by increasing buffer size as described in Section~\ref{subsec:optimizations}. First, we observe that increasing the value of $B$ for a fixed $N$ increases $\gamma_m$, i.e. more flexibility in $\epsilon$.
Secondly, an energy gain by increasing $B$ for all $\epsilon$ and a fixed $N$ is evident. It depends on the system design that which $B$ needs to be employed for a particular performance guarantee. Let us discuss the case for parameters $N=1,\theta_{\rm tar}=0.3,\epsilon=0.01$. The system with $B=0$ provides system energy of almost $-2$ dB as shown in Fig.~\ref{fig:energy_nobuffer}. If we want the same performance at reduced energy, $B=1$ provides a gain of $\Delta E=1.9$ dB. If $\Delta E>1.9$ dB, $B>1$ is required. For the same set of parameters, $B=2$ provides $\Delta E$ equal to 3.1 dB. A similar comparison can be drawn for $N=2$ and $B>0$.

A comparison of the curves for the cases $N=2,B=1$ and $N=1,B=2$ (with same $M=3$) shows that increasing DOF in any parameter $(B,N)$ is energy efficient as compared to the case $N=1,B=1$ but the effect differs widely in many ways, e.g., value of $\gamma_m$ for both cases. Similarly, increasing $B$ to reduce system energy affects system cost while increasing $N$ costs performance loss in terms of jitter.
Thus, system's energy, packet loss and latency requirements determine the parameters required to achieve performance in terms of statistical guarantee on continuity constraint.
\section{Conclusions}
\label{sect:conclusions}
We consider energy efficient multiuser scheduling over fading channels for packet loss tolerant applications. Packet loss is modeled by an average
packet drop rate and \emph{continuity constraint} on successive dropping of packets. The proposed scheme is analyzed for imperfect CSI case using
packet level channel model where the effect of imperfect CSI is modeled by fixed packet drop and success probabilities. We formulate the optimization problem for achieving minimum system energy for a target statistical guarantee on continuity constraint. Through stochastic optimization framework, we characterize the limits on achievable statistical guarantees on continuity constraint. We evaluate the effect of buffer size on the problem settings and validate numerically that buffer size can be traded to achieve better energy efficiency for a given statistical guarantee on continuity constraint and average
packet drop rate. We conclude that application's energy and latency requirements are important to determine preferable buffer size to achieve
system performance in terms of protection to bursty packet loss.

\section*{Acknowledgements}
{This work was made possible by NPRP 5 - 782 - 2 - 322 from the Qatar National Research Fund (a member of Qatar Foundation). The statements made herein are solely the responsibility of the authors.}

\bibliographystyle{IEEEtran}

\bibliography{bibliography}
\end{document}